\documentclass[manuscript]{aastex62}

\graphicspath{{./}{figures/}}

\received{January 18, 2019}
\revised{January 19, 2019}
\accepted{\today}
\submitjournal{ApJ}

\shorttitle{Spectral decomposition}
\shortauthors{Cheung et al.}

\usepackage{natbib}
\newcommand{\bart}[1]{{#1}}
\newcommand{\juan}[1]{{#1}}
\newcommand{\paola}[1]{{#1}}
\newcommand{\mcmc}[1]{{#1}}
\newcommand{\vhh}[1]{{#1}}

\begin{document}

\title{Multi-component Decomposition of Astronomical Spectra by Compressed Sensing}

\correspondingauthor{Mark C. M. Cheung}
\email{cheung@lmsal.com}

\author{Mark C. M. Cheung}
\affil{Lockheed Martin Solar \& Astrophysics Laboratory \\
3251 Hanover St, Palo Alto, CA 94304, USA}
\affil{Hansen Experimental Physics Laboratory, Stanford University \\
452 Lomita Mall, Stanford, CA 94305, USA}

\author{Bart De Pontieu}
\affil{Lockheed Martin Solar \& Astrophysics Laboratory \\
3251 Hanover St, Palo Alto, CA 94304, USA}
\affil{Rosseland Centre for Solar Physics, University of Oslo\\
P.O. Box 1029 Blindern, NO0315 Oslo, Norway}
\affil{Institute of Theoretical Astrophysics, University of Oslo\\
P.O. Box 1029 Blindern, NO0315 Oslo, Norway}

\author{Juan Mart\'inez-Sykora}
\affil{Lockheed Martin Solar \& Astrophysics Laboratory \\
3251 Hanover St, Palo Alto, CA 94304, USA}
\affil{Bay Area Environmental Research Institute \\
3251 Hanover St, Palo Alto, CA 94304, USA}

\author{Paola Testa}
\affil{Harvard-Smithsonian Center for Astrophysics \\
60 Garden St, Cambridge, MA 02193, USA}

\author{Amy R. Winebarger}
\affil{NASA Marshall Space Flight Center \\
Huntsville, AL 35812, USA}

\author{Adrian Daw}
\affil{NASA Goddard Space Flight Center \\
Greenbelt, MD 20771, USA}

\author{Viggo Hansteen}
\affil{Rosseland Centre for Solar Physics, University of Oslo\\
P.O. Box 1029 Blindern, NO0315 Oslo, Norway}
\affil{Institute of Theoretical Astrophysics, University of Oslo\\
P.O. Box 1029 Blindern, NO0315 Oslo, Norway}
\affil{Lockheed Martin Solar \& Astrophysics Laboratory \\
3251 Hanover St, Palo Alto, CA 94304, USA}

\author{Patrick Antolin}
\affil{University of St Andrews \\
St Andrews, KY16 9AJ, United Kingdom}

\author{Theodore D. Tarbell}
\affil{Lockheed Martin Solar \& Astrophysics Laboratory \\
3251 Hanover St, Palo Alto, CA 94304, USA}

\author{Jean-Pierre Wuelser}
\affil{Lockheed Martin Solar \& Astrophysics Laboratory \\
3251 Hanover St, Palo Alto, CA 94304, USA}

\author{Peter Young}
\affil{NASA Goddard Space Flight Center \\
Greenbelt, MD 20771, USA}

\author{the MUSE team}



\begin{abstract}
The signal measured by an astronomical spectrometer may be due to radiation from a multi-component mixture of plasmas with a range of physical properties (e.g. temperature, Doppler velocity). Confusion between multiple components may be exacerbated if the spectrometer sensor is illuminated by overlapping spectra dispersed from different slits, with each slit being exposed to radiation from a different portion of an extended astrophysical object. We use a compressed sensing method to robustly retrieve the different components. This method can be adopted for a variety of spectrometer configurations, including single-slit, multi-slit (e.g., the proposed MUlti-slit Solar Explorer mission; MUSE) and slot spectrometers (which produce overlappograms).
\end{abstract}



\section{Introduction} \label{sec:intro}

From solar flares to quasars, spectrometers are used to investigate a wide variety of astrophysical phenomena. To obtain measurements of the physical conditions of the emitting material (e.g. extreme UV emission lines of many-times ionized Fe atoms) inevitably requires a forward model with the following components:
\begin{itemize}
    \item \textbf{P} - A \emph{physics-based model} of the radiative process operating in the astrophysical object of interest (e.g. emission, absorption, scattering, gravitational redshift),
    \item \textbf{O} - An \emph{optical model} of the telescope (including the point-spread function, instrumental spectral broadening, and if the telescope is ground-based, atmospheric seeing), and
    \item \textbf{D} - A \emph{detector model} capturing the properties of the sensing system (e.g. non-linearity,  dark current, gain patterns, sources of noise).
\end{itemize}
\noindent The goal of spectroscopic measurements is to provide observational constraints of the physical properties of the system. For the sake of discussion, suppose one has perfect knowledge of the optical system and the detector. Even then, a physics model is still required for the most basic of spectroscopic measurements, such as the Doppler shift of a spectral line. For instance, consider spectroscopic measurements of extreme UV (EUV) emission lines from solar coronal plasma in the optically thin regime in the absence of scattering. In order to extract the Doppler shift of the line, the local enhancement (emission line) or deficit (absorption line) in the detected spectrum must first be associated with a known spectral line from a certain atomic species. This provides a reference rest wavelength of the line against which a Doppler shift (in wavelength and in velocity) can be measured. Accounting for, or in the absence of, gravitational redshift, the Doppler shift informs us about the motion of plasma along the line-of-sight (LOS), $v_{\rm LOS}$. Given an atomic model~\citep[e.g. CHIANTI,][]{Dere:1997,Young:2009,Landi:2013} we can also attribute the emission line to plasma \mcmc{in a certain range of temperatures}. Assuming thermal equilibrium conditions (e.g. thermal collisional excitation rates balanced by spontaneous radiative de-excitation), the atomic model also provides the thermal width $\sigma_{\rm th}$ of the line.

In the general case, the spectrum measured at the detector may be due to a heterogeneous mixture of plasmas at different temperatures and Doppler velocities. For example, measurement of a spectral line with an observed width $\sigma_{\rm obs} > \sigma_{\rm th}$ suggests the emitting plasma has multiple components moving at different LOS velocities (which, depending on the physics model, may be interpreted as a sign of turbulence;  \citealt{2018ApJ...864...63P,2016ApJ...820...63B,vanBallegooijen:awt}). The detection of multiple spectral lines associated with different temperatures suggests multiple thermal components in the emitting plasma. The presence of multiple components contributing to a single spectrum on the detector may be due to spatial inhomogeneities along the LOS or within the plane-of-sky area visible to the slit (or multiple slits in the case of a multi-slit instrument).

The aim of this work is to describe a method for decomposition of the spectra into constituent components of the emitting spectra by techniques of compressed sensing. The driver behind this work is to address the complexities introduced by the use of multi-slit instrumentation in solar physics (e.g., the Multi-slit Solar Explorer or MUSE, proposed as a NASA small explorer satellite), which promises to greatly increase the field-of-view and/or drastically improve the temporal cadence of spectroscopic data. \mcmc{In this paper, the method is demonstrated in the context of MUSE, for which the goal is to account for any effect of blends on the primary lines of interest, rather than to use the decomposition results directly.} However, the method described is very general and can be applied to a host of other astrophysical problems. The article is structured as follows. Section~\ref{method} gives a mathematical description of the problem for the general case. Section~\ref{sec:param} briefly discusses the parameter space we explore. Section~\ref{sec:solution} describes in some detail the compressed sensing approach, while  section~\ref{sec:examples} discusses some examples of application of the method to solar spectral data. Finally, in section~\ref{sec:discussion} we discuss the presented method and its results.

\section{Problem statement}\label{method}
\begin{figure*}
    \centering
    \includegraphics[angle=0,width=\textwidth]{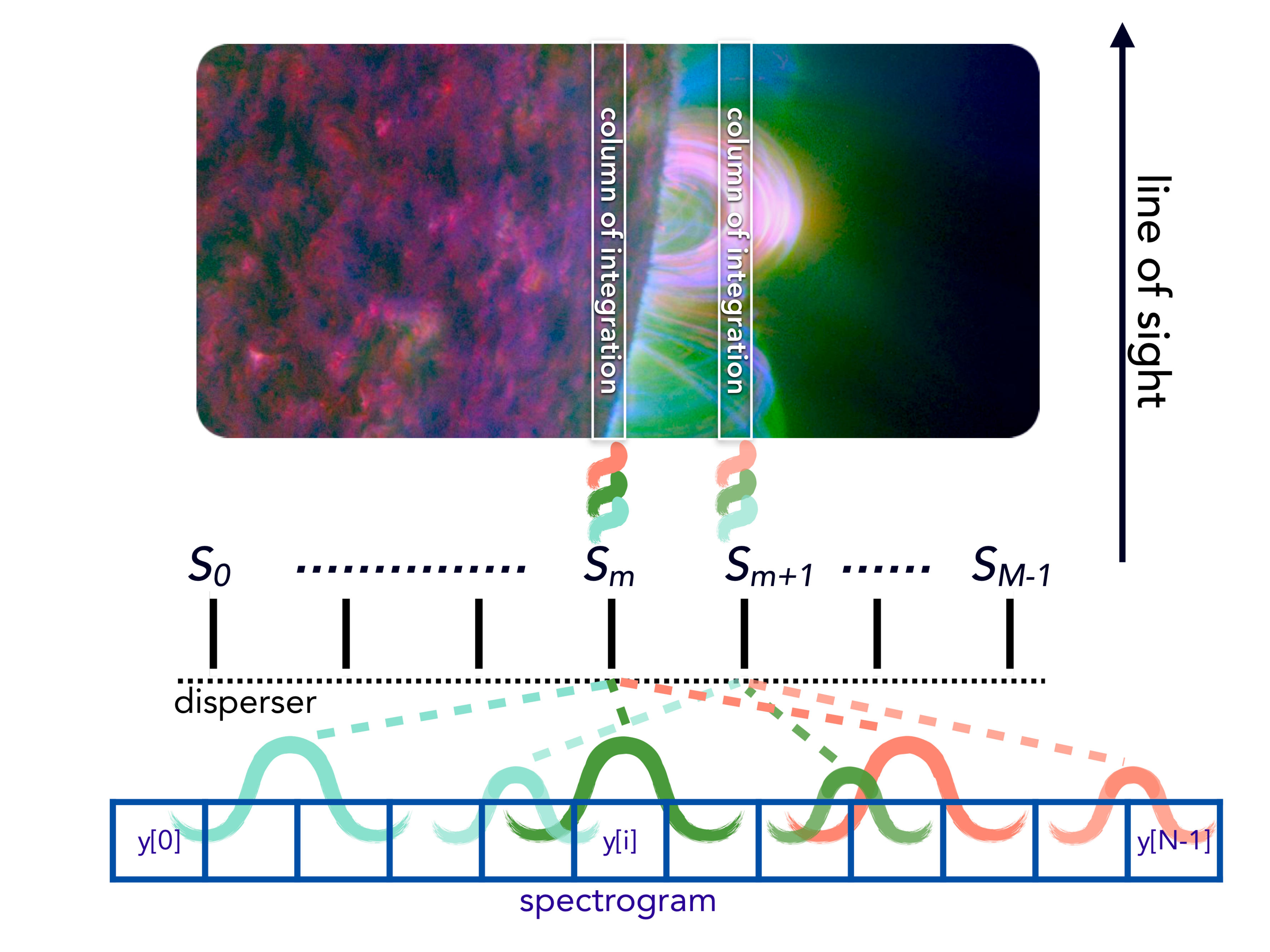}
    \caption{\bart{Schematic of a multi-slit spectral sensing system: Multi-wavelength radiation $I_\lambda$ emitted by an extended astrophysical object (e.g. the coronal plasma of a solar active region) is focused by a telescope on to a system of $M$ parallel slits $S$ that transmits light sampling a picket fence subset of the overall field of view. A disperser optical element (e.g., a diffraction grating), or system of elements, then disperses the transmitted radiation as a spectrum and focuses it on the detector consisting of an array of $N$ pixels.}
    The measured spectrogram $y[i]$ is the superposition of spectra originating from all slits.}
    \label{fig:1}
\end{figure*}

Consider a multi-slit sensing system with a set of parallel slits $\mathbf{S} = \{S_m; m=0,...,M-1\}$, each of width $w$, and a common detector at the focal plane
(see Fig.~\ref{fig:1}). \mcmc{Assume the regular slit spacing $d \gg \lambda f/A$, where $\lambda$ is the wavelength of observation and $f/A$ is the f-ratio of the spectrograph at the slit. The spectrograph disperses the light from each slit into an independent spectral intensity pattern $I_m(\lambda)$ on the detector.}

The detector has a 2D array of pixels indexed $[i, j]$, and we assume $i$ corresponds precisely to the spectral dimension and j corresponds precisely to the spatial dimension along the slit, as can be accomplished through geometric corrections when instrument alignment is not perfect. Variations along $i$ and $j$ are therefore separable, and we need only to consider variations in $i$ for the decomposition. The measured spectrogram $y[i]$ (intensity at the $i$-th pixel) can have contributions from photons originating from multiple slits.
\begin{equation}
y[i] = \mathbf{D}\left(\Sigma_{m=0}^{M-1} \mathbf{O}_m(\mathbf{P}(\Psi))[i]\right)
\label{eqn:statement}
\end{equation}
\noindent where $\mathbf{O}_m (\mathbf{P}(\Psi))[i]$ is the spectrogram due solely from radiation from slit $m$. The emitting material seen by slit $m$ has a density function $\Psi$ over some parameter space of physical properties (e.g. temperature, density and Doppler velocity) and radiates a spectrum denoted $I_\lambda = \mathbf{P}(\Psi)$. The operator $\mathbf{O}_m$ acting on $\mathbf{P}(\Psi)$ represents how the radiation is processed by the optical system, including how light through an individual slit is \mcmc{dispersed and focused} onto the detector to form a spectrum. For two adjacent identical slits, the same packet of material residing in the astrophysical object of interest with physical property $\Psi$ would lead to the same spectrum $\mathbf{O}(\mathbf{P}(\Psi))$, except $\mathbf{O}_{m+1}(\mathbf{P}(\Psi))$ would be translated from $\mathbf{O}_{m+1}(\mathbf{P}(\Psi))$ by some pixel offset $\Delta i$, namely 
\begin{equation}
    \mathbf{O}_{m+1}(\mathbf{P}(\Psi))[i] = \mathbf{O}_{m}(\mathbf{P}(\Psi))[i+\Delta i]
\end{equation}

\mcmc{For an ideal detector, the operator $\mathbf{D}$ gives an identity mapping (i.e. it preserves the spectrograph exactly). A noisy, linear detector can be described as $\mathbf{D}(\vec{y}) = G\vec{y} + \vec{e}$, where $G$ is the gain and $\vec{e}$ is the stochastic (and perhaps systematic) noise introduced by the detector.}

In general, different slits will be exposed to incident radiation from different parts of the astrophysical scenery. So the challenge is to decompose the net measured spectrum $y[i]$ into constituent components by identifying  
\begin{enumerate}
    \item the slit(s) which contributed to the net spectrogram, and
    \item the physical properties of the radiating material along the column of integration in the LOS seen by each slit. 
\end{enumerate}
\noindent Though the above description of the problem applies to both radiation from optically thick and optically thin plasmas, the rest of this paper will be concerned with the simple case of optically thin plasmas. This allows us to express the physical model $\mathbf{P}$ as a linear operator over the parameter space density function $\Psi$. \textbf{The aim of \emph{measuring} the physical properties of emitting material is equivalent to finding the function $\Psi$ such that Eq.~(\ref{eqn:statement}) is satisfied.} The following sections (in particular section~\ref{sec:examples}) will provide some concrete examples.

\section{parameter spaces} \label{sec:param}
\subsection{Differential Emission Measure (DEM)}\label{sec:dems}
A common concept encountered in EUV and X-ray observations of solar plasma is the \emph{Differential Emission Measure}~\citep[see, e.g.][]{Boerner:AIA}, defined by the relation
\begin{equation}
    y[i] = \int_{T_0}^{T_1} K_i(T) {\rm DEM(T)}dT,
\end{equation}
\noindent where $y[i]$ is the detected spectrogram (after dark subtraction, flatfielding, correction for nonlinearities, etc), $K_i(T)$ is the temperature response function of $i-$th spectral channel, and DEM$(T)dT = \int_0^\infty n_e(T)^2 dl$ is the electron density squared, integrated along the line-of-sight, contained in a temperature bin of width $dT$. $K_i(T)$ encapsulates assumptions about the radiative properties of the emitting plasma (i.e. $\mathbf{P}$) and the optical properties of the system (i.e. $\mathbf{O}$), such as \mcmc{point spread} function, effective area etc. 

\mcmc{The aim of the DEM inversion problem in previous work was to recover DEM(T) given a set of measurements} $y[i]$~\citep[see][and references therein]{Cheung:AIADEMs}. In that context, DEM$(T)$ is the parameter space density function $\Psi$, which spans over the temperature range $[T_0,T_1]$\mcmc{, but did not include the dimensions of Doppler velocity, as we assumed that the observing system in question}~\citep[e.g. the EUV channels of the Atmospheric Imaging Assembly see][]{Lemen:AIA,Boerner:AIA} does not have sufficient spectral resolution to resolve Doppler shifts. Furthermore, the physical model of radiation assumes the emissivity of EUV spectral lines is only a function of $T$~\citep[though as shown by][there is also a slight dependence on plasma density]{Martinez-Sykora:2011vn,Testa:DEMs}.

\subsection{Velocity DEM (VDEM)}\label{sec:vdems}
As an extension of the DEM inversion problem, consider the situation where the sensing system has sufficient spectral resolution and sampling to be sensitive to Doppler shifts. In this case the parameter space density function $\Psi(T,v)$ spans temperature and Doppler velocity space. Solving the VDEM problem means we seek to quantify how much plasma (in terms of emission measure $n_e^2$) is at a certain temperature $T$ moving with Doppler velocity $v$. In other words, find $\Psi(T,v)$ such that the following relation holds:

\begin{equation}
    y[i] = \int_{v_0}^{v_1}\int_{T_0}^{T_1} K_i(T,v) \Psi(T,v) dT dv.\label{eqn:vdem}
\end{equation}

In the case of multiple slits,  the r.h.s. of Eq.~(\ref{eqn:vdem}) also includes a sum over $m$ (the index for the M slits) and $K_{i,m}$ may be slit-dependent. In section~\ref{sec:solution}, we describe how a compressed sensing technique is used for the class of problems similar to Eq. (\ref{eqn:vdem}).

\section{Compressed Sensing Solution Approach}
\label{sec:solution}
Suppose $\Psi\ge 0$ is a density function over an $n$-dimensional parameter space $\vec{\phi} = (\phi_0, \phi_1, \phi_j, ... , \phi_{n-1})$ where $\phi_j\in [\phi_{j,0},\phi_{j,1}]$, and $\phi_{j,0}$ and $\phi_{j,1}$ are the lower and upper bounds of the $j$-th parameter. The operator $\mathbf{P}$ acts on $\vec{\phi}$ to generate (and propagate) radiation to the sensing system. The operator $\mathbf{O}$ takes this incident radiation arriving at the sensing system (e.g. a telescope and its optical system) and produces the $M$-tuple $\vec{y}$ (each component of $\vec{y}$ is the spectrogram measured in a pixel of the detector). $\mathbf{P}$ and $\mathbf{O}$,~\mcmc{and the deterministic part of $\mathbf{D}$ (e.g. gain)} are assumed known.

The solution strategy begins with generating response functions. Consider each dimension of parameter space is discretized into a finite number of points $N_j$. For each point $\vec{\phi}$ in parameter space, compute the detector response $\vec{r}_{\vec{\phi}}$ according to Eq.~(\ref{eqn:statement}). The set of all response vectors are used to generate the response matrix $\mathbf{R} = \left (\vec{r}_{\vec{\phi}}\right)$. $\mathbf{R}$ has dimensions $N\times M$, where $N = \Pi_{j=0}^{n-1} N_j$. Parameter estimation (i.e. measuring the physical properties of the radiating plasma) is then equivalent to solving the following linear system for $x$
\begin{equation}
    \vec{y}=\mathbf{R}\vec{x},\label{eqn:linearsystem}
\end{equation}
\noindent where $\vec{x} = (x_j), x_j \ge 0 $ is an $N$-tuple of coefficients for the response functions. In other words, we seek to express the measured spectrogram $\vec{y}$ as a linear superposition of response functions $\vec{r}_{\phi}$. For a multi-dimensional parameter space, this linear system may be underdetermined. There are a variety of compressed sensing schemes that can be used to tackle such types of problems. For instance, for  DEM reconstruction from narrowband EUV data taken by the Atmospheric Imaging Assembly~\citep[AIA;][]{Lemen:AIA, Boerner:AIA} onboard NASA's Solar Dynamics Observatory~\citep[SDO;][]{Pesnell:SDO},~\citet{Cheung:AIADEMs} presented a validated inversion scheme based on basis pursuit~\citep{Chen:AtomicDecomposition}. That particular problem has $M=6$ (six EUV channels used) and $N\approx 20$ (number of $\log T$ bins). For much larger problem sizes, we found the lasso method~\citep{Tibshirani:lasso} implemented in the Python scikit learn module~\citep{scikit-learn} to give reliable results. Lasso seeks a solution for Eq. (\ref{eqn:linearsystem}) by finding:

\begin{equation}
    \vec{x}^{\#} = {\rm argmin }\left[ \frac{1}{2}\left( \vec{y}-\mathbf{R}\vec{x}\right)^2 + \alpha |\vec{x}|_1\right], \label{eqn:lasso}
\end{equation}
\noindent where $|\vec{x}|_1 = \Sigma_{i=0}^{N-1} |x_i|$ is the L1 norm of $\vec{x}$. In other words, $\vec{x}^{\#}$ is the argument $\vec{x}$ which minimizes the objective function in the square brackets. By adding a L1 penalty term, Lasso promotes sparsity in the solution.~$\alpha$ is a hyperparameter (i.e. a parameter that is not fitted by the algorithm) used to control the level of sparsity. \mcmc{A larger value of $\alpha$ tends to yield solutions that have smaller L1 norm (more sparse).}

\section{Examples of spectral sensing systems}
\label{sec:examples}

\subsection{Single-slit spectrometer: Hinode/EIS}\label{singleslit}
Perhaps the most common type of astronomical spectrometer instruments are those with a single slit,~i.e.~$M=1$. An example is the EUV Imaging Spectrometer~\citep[EIS][]{Hinode:EIS} on board the Hinode mission~\citep{Kosugi:Hinode}. Hinode/EIS is sensitive to emission lines of many-times ionized metallic species (e.g. Fe, O and Ni) found in solar coronal plasmas at temperatures between $\log T/{\rm K} \sim 5.0$ and $\log T/{\rm K} \sim 7.3$.

\begin{figure*}
\centering
\includegraphics[width=0.77\textwidth]{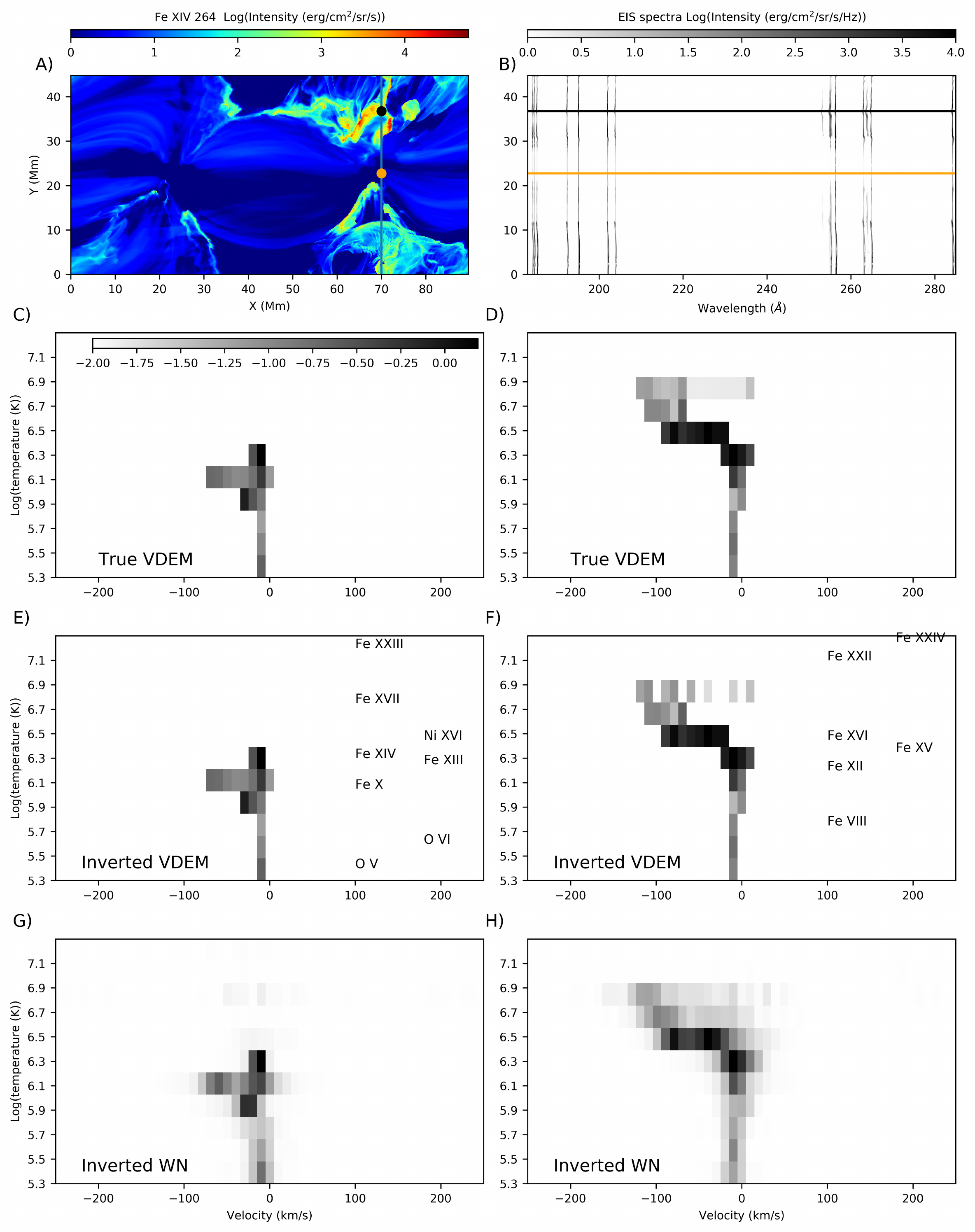}
\caption{Example of VDEM inversion for Hinode/EIS spectrograms. (A) Synthesized spatial intensity image of the \ion{Fe}{16} $264$~\AA~line for a snapshot from radiative MHD simulation of a solar flare~\citep{CheungRempel:2018}. (B)  the spectrogram if the EIS slit were placed along the vertical line ($x=70$ Mm) in panel (A). The ground truth VDEMs (i.e.~as sampled from the MHD simulation) at two positions (in order of increasing $y$ coordinate position) are shown in panels (C) and \paola{(D)}, respectively. The corresponding recovered VDEMs \juan{are shown in \paola{(E)} and (F) (for the case without photon noise) and in (G) and (H) (averaged over 250 realizations of photon noise)}, both for the locations marked in yellow and black in panels (A) and (B). The colorbar inset in panel (C) shows the emission measure in units of $10^{27}$ cm$^{-5}$. \paola{The ions emitting the lines used for the VDEMS inversions are labeled in panels (E) and (F), and their labels are positioned at the temperature corresponding to the peak of the line emissivity function ($G_n(T)$).}}\label{fig:eis}
\end{figure*}

In general, an EIS spectrogram consists of multiple emission lines corresponding to different plasma components of various temperature and Doppler velocities. Sample spectrograms are shown in Fig.~\ref{fig:eis}. Using a snapshot from a three-dimensional radiative MHD simulation of a solar flare~\citep{CheungRempel:2018}, we synthesized the optically thin emergent EUV radiation using atomic models from the CHIANTI database ~\citep[version 8][]{DelZanna:2015}. A spatial intensity image of the \ion{Fe}{16} $264$~\AA~line is shown in panel (A) of Fig.~\ref{fig:eis}. Panel (B) of the same figure shows the spectrogram if the EIS slit were placed along the vertical line ($x=70$ Mm) in panel (A). Along the EIS slit, we sample VDEMs from two positions indicated by the colored dots in panel (A) and the corresponding colored horizontal lines in panel (B). The ground truth VDEMs (i.e., as sampled from the MHD simulation) at these two positions (in order of increasing $y$ coordinate position) are shown in panels (C) and (D) respectively.

To invert the spectrograms for VDEMs, we follow the procedure outlined in section~\ref{sec:solution}. Consider a unit of emission measure (e.g. for observations of solar plasmas at 1 AU, an emission measure of $EM_0 = 10^{25}$ cm$^{-5}$ or above is generally detectable given a sufficiently bright emission line and sufficient effective area and exposure time). Consider a VDEM distribution $\Psi(T,v) = EM_0\delta(T_0,v_0)$, i.e. \bart{an isothermal plasma at temperature $T_0$ at a single line-of-sight velocity $v_0$.} 
Using the physical model $\mathbf{P}$ (CHIANTI) and instrumental response model ($\mathbf{O}$), compute the response vector $\vec{r}(T_0,v_0)$ for this VDEM distribution, and repeat for other values of $T_0$ and $v_0$ in VDEM space. This allows us to construct the response matrix $\mathbf{R}$ used for VDEM inversions. The parameter space has a velocity range of $\pm400$~km~s$^{-1}$ with a velocity bin size of 10~km~s$^{-1}$ and the temperature ranges from $\log (T[{\rm K}]) = 5.3$ to  $\log (T[{\rm K}]) = 7.3$ with a bin size of 0.2. i.e., $N=80\times9=720$. For the radiation model (i.e.~$\mathbf{P}$) we consider plasma with solar coronal abundances. The emission model includes 18 spectral lines (from \citealt{Doschek:2013ce}) which are listed in panels (E) and (F) of Fig.~\ref{fig:eis}, and cover a broad temperature range from $\log (T_{\rm max}[{\rm K}]) \sim 5.45$ (\ion{O}{5}~248.46\AA), to $\log (T_{\rm max}[{\rm K}]) \sim 7.2$ (\ion{Fe}{24} 255.1\AA).  
\bart{The detector model $\mathbf{D}$ includes photon noise, based on Poisson distribution and exposure times of 30 s, using the EIS effective area.}
Panels (E) and (F) show the VDEM distributions inverted from the spectrogram displayed in panel (B) 
\juan{, while panels (G) and (H) show the average of the inverted VDEMs over 250 realizations of the photon noise}. Although there are imperfections in the recovered VDEM, they reproduce the salient features in the ground truth VDEMs.
Over a range of temperatures, the inversion correctly reproduces the spread of emission measure in Doppler velocity space. For instance, the range of Doppler velocities with significant emission measure is much narrower for $\log T[{\rm K}]<6$. 

\subsection{Multi-slit spectrometer}\label{multislit}
The MUlti-Slit Solar Explorer (MUSE) is a science mission proposed to NASA's Small Explorer program
\bart{\cite{2017SPD....4811008T}}. Like Hinode/EIS, it measures atomic EUV lines emitted by coronal plasma. However, by using 37 spectrally dispersing slits (i.e.~$M=37$), it allows spatial rasters of solar active regions up to two orders of magnitude faster than EIS or any other existing or planned spectrometers. This design allows the MUSE instrument to capture many more solar eruptions and flares, and, for the first time, capture them with sufficient spatio-temporal resolution to reveal the dynamic evolution of the active corona. The extremely rapid (12s cadence), sub-arcsecond (0.4 arcsec) resolution rasters (170 x 170 arcsec$^2$) with broad temperature coverage, accompanied by large FOV context imaging in several EUV lines (\ion{Fe}{12}~195~\AA~and \ion{He}{2}~304~\AA) will allow MUSE to address its top-level science goals: 1. determine which mechanisms drive coronal heating and the solar wind, 2. understand the genesis and evolution of the unstable solar atmosphere, and 3. investigate fundamental physical processes in the solar corona. The three spectral passbands are dominated by  spectral lines with wavelengths around 108\AA (\ion{Fe}{19}, \ion{Fe}{21}), 171\AA\ (\ion{Fe}{9}) and 284\AA\ (\ion{Fe}{14}). These lines are formed around $\log T[{\rm K}] = 7.0,~7.1,~5.7,~6.4$, respectively. Because the passbands are spectrally wider than the (wavelength) separation between neighboring slits, the multi-slit design can, in principle, lead to overlap of spectral information from neighboring slits. This is minimized by: \bart{1. the selection of band passes to study bright, well-isolated lines as primary diagnostics, 2. the selection of a slit spacing that minimize possible blends from other slits.} This typically limits multi-slit confusion to regions in which the primary lines are not bright, or where the plasma has unusual emission measure distributions (e.g., a predominance of very cool plasma, e.g., in coronal loop fans, which can lead to contamination by secondary lines). Our spectral decomposition code has been shown to be very effective in disambiguating the multi-slit confusion \vhh{even} for these \vhh{difficult} conditions (as shown below, \bart{and in more detail, in a follow-up paper focusing on MUSE applications}).

To satisfy the science requirements for MUSE, it is \bart{not} necessary to accurately determine a VDEM distribution for each slit S:  $\Psi(T,v,S)$. Instead this VDEMS distribution is only used as an intermediate step to disambiguate any multi-slit confusion of the primary lines, e.g., by calculating, for each slit, the primary lines and \bart{secondary} lines from the VDEMS distribution. Similar to the example in \S~\ref{singleslit}, we use the physical model $\mathbf{P}$ (CHIANTI) and instrumental response model ($\mathbf{O}$, which takes into account the position of the 37 slits in the spectrogram) to compute the response vector $\vec{r}(T_0,v_0, S_0)$ for this VDEMS distribution, and repeat for other values of $T_0$ and $v_0$ in VDEM space and the 37 slits $S_0$. This allows us to construct the response matrix $\mathbf{R}$ used for VDEMS inversions. The matrix $\mathbf{R}$ has a velocity range of $\pm400$~km~s$^{-1}$ with a velocity bin size of 10~km~s$^{-1}$, a temperature ranges from $\log T[{\rm K}] = 4.65$ to  $\log T[{\rm K}] = 7.85$ with a bin size of 0.2, and is calculated for all 37 slits. For the detector model $\mathbf{D}$, we added photon noise based on a Poisson distribution, with exposure times of 1.5s and the MUSE effective area.

\begin{figure*}
    \centering
    \includegraphics[width=0.98\textwidth]{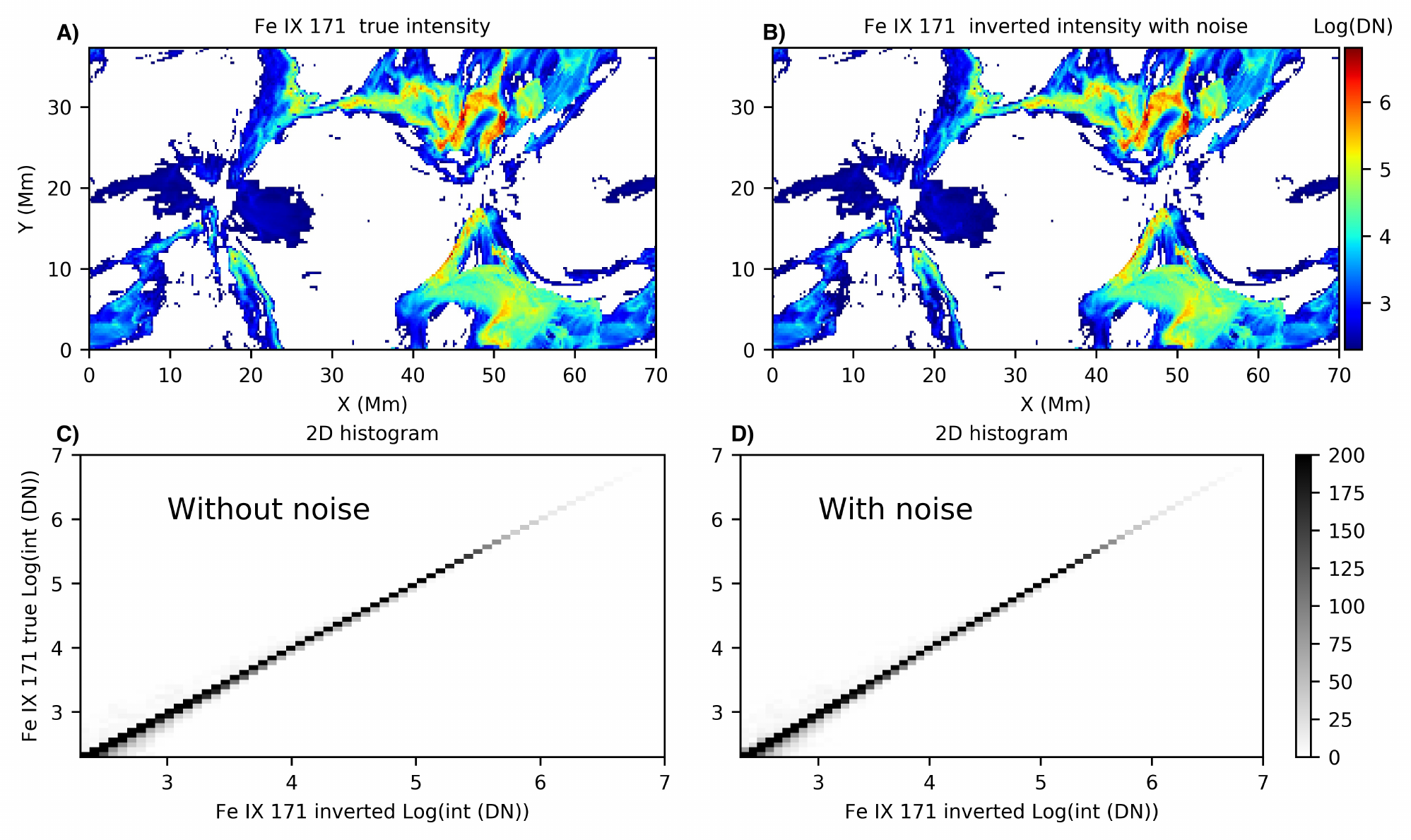}
    \caption{Ground truth intensity map of \ion{Fe}{9}~171~\AA\ from the same snapshot shown in Figure~\ref{fig:eis} \juan{(panel A)} is nicely reproduced by the synthetic \ion{Fe}{9}~171~\AA\ intensity \juan{(panel B)} calculated after applying the spectral decomposition code on synthetic multi-slit MUSE observations \juan{including photon noise}. The good correlation is illustrated by the right \juan{panels} which shows \juan{2D histograms} of the ground truth (vertical axis) and inverted intensities (horizontal axis) \juan{without and with photon noise in panels (C) and (D), respectively.}}
    \label{fig:muse}
\end{figure*}

Using the same snapshot of the 3D radiative MHD simulated solar flare from the previous section, we synthesized the optically thin MUSE spectrum using the CHIANTI database for all three passbands (108~\AA, 171~\AA\ and 284~\AA, i.e. N=600). The synthetic MUSE observations take into account the instrument response for the 37 slits, the spectral and spatial resolution, and \mcmc{all spectral lines from the CHIANTI database with wavelengths that could fall on the detector from any slit. Specifically, spectra were generated using all lines from CHIANTI v8.0.7, with updated data for \ion{Fe}{7}}. We use the spectral decomposition code on the synthetic MUSE data from all three passbands to determine the VDEMS. As mentioned, this inverted VDEMS is not the end goal, but instead can be used to reconstruct the dominant spectral lines without contribution from other non-dominant spectral lines and removing the confusion from adjacent slits. The example in Figure~\ref{fig:muse} compares the synthetic \ion{Fe}{9}~171~\AA\ intensity map from the 3D radiative MHD flare simulation (ground truth, \juan{ panel (A)}) with a disambiguated map from the inverted VDEMS \bart{(based on synthetic data that includes photon noise, panel (B))}. The correlation between the ground truth and intensities derived from the inversions are very good (\bart{both for the case without photon noise, panel (C), and with photon noise, panel (D)}), illustrating that the spectral decomposition code accurately disambiguates the MUSE data, even for the challenging scene presented by a flare (in which secondary lines and strong Doppler shifts can potentially lead to multi-slit confusion). Note that this is only shown as an illustration of how the decomposition code can disambiguate multi-slit spectra: in the concept of operations of MUSE, the disambiguation code would be used to identify locations of multi-slit confusion so that users can isolate signals from secondary lines or neighboring slit before analyzing the primary lines. \bart{For the application of this method to MUSE data, a more detailed description of the various trade-offs and optimal choices for velocity bins, temperature bins and inversion parameters, as well as the dependence on abundance, density and noise, will be provided in a follow-up paper \citep{Martinez-Sykora2019}.}

\subsection{\mcmc{Slot and slitless spectrometers}}\label{slot}
\mcmc{A special case is when the slit spacing $d$ is equal to the slit width $w$ (in terms of angular coverage over the plane-of-sky). This allows us to treat slitless spectrometers and spectrometers with slot modes within this multi-slit framework. An example of a slitless spectrometer is the S082A instrument on Skylab~\citep{Tousey:1977}. The Coronal Diagnostic Spectrometer (CDS) onboard the Solar and Heliospheric Observatory and the Hinode/EIS instrument~\citep{Hinode:EIS} are spectrographs with slot modes}. The validation of this method to decomposing spectra from these types of instruments like the proposed COronal Spectroscopic Imager in the EUV (COSIE) is detailed in a companion paper~\citep{Winebarger:COSIE}.

\section{Discussion} \label{sec:discussion}
In this paper, we outlined a general approach to performing the decomposition of spectrograms from astronomical spectrographs with a variety of configurations (e.g. single-slit, multi-slit and slot mode). The decomposition of spectrograms is not only helpful for removing possible ambiguities (e.g. from which slit did this detector signal originate). Such decomposition is \bart{also useful} for estimating the physical parameters of interest. In this paper and in the companion paper~\citet[][ on decomposing overlappograms from slot spectrometers]{Winebarger:COSIE}, we demonstrate application of the technique for interpreting spectrograms from different instruments observing the solar corona.

The applications considered in this paper were for EUV radiation emanating from optically thin plasmas. However, the method is also useful for the interpretation and inversion of spectra from optically thick plasmas. Suppose the spectrometer is designed to detect an atomic absorption line formed in the (partially) optically thick photospheres or chromospheres of astrophysical objects. The emergent spectrum from the optically thick material depends on a number of physical parameters describing the atmosphere including the number density of the absorbing species, the ambient temperature, the slope of the source function $S_\nu(\tau)$ as a function of optical depth $\tau$, the local magnetic field (if magneto-optical effects like Zeeman splitting are important) and more (especially if the line does not form in local thermodynamic equilibrium). Nevertheless, given a physics model including the desired effects, one can construct a library of emergence spectra over parameter space, and fold them through the optical model $\mathbf{O}$ to compute the response matrix $\mathbf{R}$. Seeking a solution along the lines of Eq.~(\ref{eqn:lasso}) remains an attempt to express the measured spectrogram as the linear superposition of spectra from the library of atmospheric models. However, since the operator $\mathbf{P}$ is not linear for optically thick radiation, the components of the solution vector $\vec{x}$ cannot be interpreted as a density of emitting material (in the case of optically thin EUV radiation, this density is the emission measure) along a single ray. Instead, the linearity should be attributed to the optical characteristics of the instrument (i.e. the operator $\mathbf{O}$).

\mcmc{Within the framework presented here, a single component inversion is equivalent to seeking to describe the spectrogram $\vec{y}$ with a single response vector $\vec{r}(\vec{\phi})$. Unless one is certain the object is spatially resolved given the telescope PSF, there is perhaps no compelling justification (other than computational expediency) for a single component inversion.} \mcmc{For instance, a spatially extended point spread function (PSF) would lead to the addition of signal associated with photons from different parts of an astrophysical object.}

The multi-slit configuration of an instrument like MUSE can be thought of as having a spatially extended (over multiple slits) PSF. But even in the absence of multiple slits, the PSF of a telescope can be sufficiently extended that variations of the physical parameters of interest in the plane-of-sky are not resolved. In such cases, the inversion of an observed spectrogram with a single atmospheric model (i.e. single set of physical parameters) may yield systematic errors. It is then perhaps more appropriate to fit the spectrogram with a linear combination of atmospheric models. Whether such an approach yields superior results over single component inversion remains to be tested and validated. The metrics of interest would depend on the specific use case and are outside the scope of this paper.

In this article, we outlined a novel framework for multi-component decomposition of astronomical spectra. We have illustrated application of the method to solar observing instruments, but it can also be used for the interpretation of spectra from other astrophysical sources.




\acknowledgments
\mcmc{MCMC, BDP, JMS and PT acknowledge support by NASA's Heliophysics Grand Challenges Research grant \emph{Physics and Diagnostics of the Drivers of Solar Eruptions} (NNX14AI14G to LMSAL).} P.A. acknowledges funding from his STFC Ernest Rutherford Fellowship (grant agreement No. ST/R004285/1).

%







\bibliography{refs,refs2,refs3}



\end{document}